 \documentclass[final,12pt]{elsarticle}

\usepackage{amssymb}
\usepackage{lipsum}
\usepackage{amsthm}

\usepackage{amsfonts}
\usepackage{mathrsfs}
\usepackage{systeme, mathtools}
\usepackage{stmaryrd}

\usepackage{tikz-cd}
\usepackage{xcolor}
\usepackage{graphicx}
\usepackage{float}
\usepackage{dcolumn}
\usepackage{bm}
\usepackage{appendix}
\usepackage{url}
\usepackage{hyperref}
\hypersetup{hidelinks}

\numberwithin{equation}{section}
\newtheorem{theorem}{Theorem}[section]

\theoremstyle{definition}

\newtheorem{definition}[theorem]{Definition}

\newcommand{\R}{\mathbb{R}}

\newcommand{\su}[1]{\mathfrak{su}(#1)}
\newcommand{\so}[1]{\mathfrak{so}(#1)}
\newcommand{\M}{\mathcal{M}}
\newcommand{\T}{\mathrm{T}}
\newcommand{\dV}{\mathrm{Vol}_q}

\newcommand{\OF}{P^{SO}(\Sigma)}
\newcommand{\SP}{P^{Spin}(\Sigma)}
\newcommand{\ad}{\operatorname{ad}}
\newcommand{\Ad}{\operatorname{Ad}}
\newcommand{\tensor}{\otimes}
\newcommand{\bra}{\langle}
\newcommand{\ket}{\rangle}

\newcommand{\imm}{\hookrightarrow}
\newcommand{\Diff}{\mathrm{Diff}(\Sigma)}

\journal{Journal of Geometry and Physics}

\begin{document}

\begin{frontmatter}



\title{Fiber bundle structure in Ashtekar-Barbero-Immirzi formulation of General Relativity}


\author[Sap,INFN]{Matteo Bruno}
\ead{matteo.bruno@uniroma1.it}

\affiliation[Sap]{organization={Physics Department, Sapienza University of Rome},
            addressline={P.za Aldo Moro 5}, 
            city={Rome},
            postcode={00185}, 
            state={},
            country={Italy}}
\affiliation[INFN]{organization={INFN, Sezione               di Roma 1},
            addressline={P.le Aldo Moro 2}, 
            city={Rome},
            postcode={00185}, 
            state={},
            country={Italy}}

\begin{abstract}
We aim to provide a rigorous geometric framework for the Ashtekar-Barbero-Immirzi formulation of General Relativity. As the starting point of this formulation consists in recasting General Relativity as an \( SU(2) \) gauge theory, it naturally lends itself to interpretation within the theory of principal bundles. The foundation of our framework is the spin structure, which connects the principal \( SU(2) \)-bundle construction with the Riemannian framework. The existence of the spin structure enlightens the geometric properties of the Ashtekar-Barbero-Immirzi-Sen connection and the topological characteristics of the manifold. Within this framework, we are able to express the constraints of the physical theory in a coordinate-free way, using vector-valued forms that acquire a clear geometric interpretation.\\

Using these geometric concepts, we analyze the phase space of the theory and discuss the implementation of symmetries through the automorphism group of the principal \( SU(2) \)-bundle. In particular we demonstrate that the description of the kinematical constraints as vector-valued forms provides a natural implementation as momentum maps for the automorphism group action.

\end{abstract}




\end{frontmatter}




\section{Introduction}
The introduction of Ashtekar variables \cite{ashtekar86,ashtekar87} has significantly transformed the Hamiltonian formulation of General Relativity \cite{ADMdyn,ADMvar}. With these variables, the theory of gravitation acquires characteristics similar to those of Yang-Mills theory \cite{Barbero1995, Immirzi_1997}, thereby necessitating the use of gauge theory language. Starting from this classical formulation, a quantum theory of gravity emerges \cite{rovelli1991ashtekar,thiemann1998quantumI,thiemann1998quantumII,thiemann1998quantumIII}. Known as Loop Quantum Gravity, this theory provides a non-perturbative quantization with interesting algebraic features \cite{RovelliKnot, RovelliNet}. Furthermore, it addresses one of the most significant aspects of quantum gravity: the quantization of geometry, manifested in the discrete spectra of geometrical operators such as area and volume \cite{ROVELLI1995}.\\
In this paper, we will rigorously demonstrate the role of the fiber bundle structure in this formulation and elucidate the connections between physical quantities and geometric objects. Despite the existence of a rigorous mathematical literature on the argument \cite{thiemann_2007,Fleischhack2015,Fatibene_2024}, a deep analysis and comprehensive composition of the Ashtekar-Barbero-Immirzi formulation within a proper geometric framework is still lacking. In particular, a unified description of the Ashtekar variables and constraints as geometric objects, rigorously defined within the context of principal bundles, along with a geometric characterization of the reconstruction process, remains absent. We aim to provide this analysis and establish a shared formal language for mathematical and theoretical physicists interested in General Relativity and Quantum Gravity. We intend to specify this framework using differential topology via principal bundle theory and extend this analysis to the constraints of Hamiltonian theory. This approach also allows us to analyze the phase space within this context. Furthermore, this study highlights the distinct differences between the structure and geometric interpretation of Yang-Mills theory and the Ashtekar-Barbero-Immirzi formulation when both are formulated within the same framework.\\

The primary objective of this manuscript is to provide a framework within fiber bundle theory for giving a geometric interpretation to the Ashtekar variables. This framework incorporates the principal bundle structure, analogous to Yang-Mills theory, but with an additional component: a spin structure. Within this framework, the pair of Ashtekar variables is interpreted in the context of connections and sections in the adjoint bundle, while the spin structure is necessary for the geometric interpretation of the reconstruction. Furthermore, we can express the constraints of the theory in a coordinate-free manner and interpret them as sections of suitable vector bundles. From a purely mathematical perspective, we will precisely define and prove that, on a 3-dimensional spin Riemannian manifold $(\Sigma, q)$, the following two sets of data are equivalent:
\begin{table}[H]
    \centering
    \begin{tabular}{c|c}
        $\omega$ & $\mathrm{K}$ \\ 
        connection in the principal $\mathrm{Spin}(3)$-bundle & $(0,2)$-symmetric tensor on $\T\Sigma$ \\ [1ex]
         & \\
        \textit{together with equations} & \textit{together with equations} \\
         & \\
        $\boldsymbol{G}(\omega)=0$ &  \\
        Gauss constraint & \\ [1ex]
        $\boldsymbol{V}(\omega)=0$ & $\boldsymbol{D}(\mathrm{K})=0$ \\
        Vector constraint & Codazzi equation\\ [1ex]
        $S(\omega)=0$ & $ H(\mathrm{K})=0$ \\
        Scalar constraint & Gauss equation
    \end{tabular}
    \caption{The two datasets are shown. On the left, the data relate to the Ashtekar-Barbero-Immirzi formulation of General Relativity. On the right, along with the metric $q$, the data pertain to the ADM formulation.}
    \label{tab:my_label}
\end{table}
The data set on the left originates from the Ashtekar-Barbero-Immirzi formulation of General Relativity. The connection $\omega$ will be associated with the Ashtekar-Barbero-Immirzi-Sen connection. We will show that the quantities $\boldsymbol{G}$, $\boldsymbol{V}$, and $S$ correspond to the constraints in this formulation. Specifically, the Gauss constraint $\boldsymbol{G}$ will be a 3-form on $\Sigma$ with values in the adjoint bundle of the spin bundle, analogous to the Yang-Mills theory \cite{KORI_2019}. The Vector constraint $\boldsymbol{V}$ will be a 3-form with values in the cotangent bundle of $\Sigma$, encoding the torsion of $\omega$. Finally, the Scalar constraint $S$ will be a 3-form related to the scalar curvature of $\omega$.\\
On the right-hand side, the data correspond to the Arnowitt-Deser-Misner (ADM) formulation of General Relativity. The supermomentum $\boldsymbol{D}$ and the superhamiltonian $H$ are a 1-form and a scalar function on $\Sigma$, respectively. These constraints are well-known equations in the context of Riemannian submanifolds; specifically, they are the contracted Codazzi equation and the Gauss equation for the scalar curvature. Using these constraints, we can interpret $\mathrm{K}$ as the extrinsic curvature of an isometric embedding of $\Sigma$ into a Ricci-flat spacetime $\M$ \cite{Choquet}. Namely, a spacetime that satisfies the Einstein Field Equation for a vacuum universe.\\

Afterward, we will discuss the implementation of the symmetries of the physical theory and the formulation without a metric within the provided framework, thereby recovering background independence to better align with the principles of General Relativity. Furthermore, we will demonstrate that these symmetries are associated with the action of the automorphism group of the spin bundle.\\
Finally, we can construct and analyze the phase space, examining its topological and symplectic properties. Using the tools of symplectic geometry, we will discuss the action of the automorphism group, providing a momentum map for this action constructed from the Gauss and Vector constraints, and demonstrating, within this framework, the well-known correspondence between the Ashtekar variables and the ADM phase space.

\section{Brief review of the physical notation}
\label{SecADM}
To construct a Hamiltonian theory of General Relativity it is necessary to introduce the Arnowitt-Deser-Misner (ADM) formalism \cite{ADMvar}. This formulation requires a globally hyperbolic spacetime, specifically a time-orientable Lorentzian manifold that admits a Cauchy surface \cite{hawking_ellis_1973}. Let $\M$ be a globally hyperbolic spacetime, Geroch's theorem \cite{geroch1970domain} ensures that $\M$ is homeomorphic to $\M\cong \R\times \Sigma$, where $\Sigma$ is a three-dimensional manifold. In particular, there exists a surjective function $f:\M \to \R$ such that $\Sigma_t\doteq f^{-1}(t)$ is a Cauchy surface homeomorphic to $\Sigma,\, \forall t \in \R$. The theorem remains valid in the smooth category, where all structures, such as manifolds and functions, are smooth \cite{Minguzzi}.\\
The collection $\mathscr{F}\doteq \{\Sigma_t\}_{t\in\R}$ clearly defines a foliation for $\M$. Indeed, for each point $p\in\M$ there exists a system of local coordinates $\{x^{\mu}\}$ such that each $\Sigma_t$ is described locally by the equation $x^0=const$.\\
Considering the splitting $\M= \R\times\Sigma$, we can give coordinates at $p=(t,x)$ as $x^0=x^0(t)$, while $x^a=x^a(x)$ coordinates on $\Sigma$. Moreover,  $\T\M\cong \T\R\times\T\Sigma$, so, on every point, this choice induces a basis of tangent vectors compatible with the splitting $\T_p\M\cong\T_{t}\R\oplus \T_x\Sigma$, in particular $\partial_{x^0}|_{t}\in\T_{t}\R$ and $\partial_{x^a}|_x\in \T_x\Sigma$.\\
In this case, the function of Geroch's theorem is the projector to the first factor $f\equiv \operatorname{pr}_1$. This function determines the vector field $n\doteq df^{\sharp}/||df||$, where $\sharp:\T^*\M \to \T\M$ is the musical isomorphism \cite{Lee2012} and $||\cdot||$ is the norm induced by the metric $g$. This vector field is, for each fixed $t\in \R$, the normal versor of $\Sigma_t$. From this, we can write the following linear combination \[\partial_{0}=Nn+N^a\partial_a.\] The functions $N(t,x)$ and $N^a(t,x)$ are called lapse function and shift vector respectively. These functions, together with the induced Riemannian metrics $q$ on $\Sigma_t$ (we omit the dependence of $q$ on $t$), characterize the metric $g$ at each point of $\M$
\begin{equation*}
    g_{ab}=q_{ab},\ \ \ g_{0a}=N^bq_{ab},\ \ \ g_{00}=-N^2+N^aN^bq_{ab}.
\end{equation*}
The other geometric quantity that describes the geometry of the hypersurface $\Sigma_t$, together with $q$, is the extrinsic curvature $\mathrm{K}$.\\
Starting from the Einstein-Hilbert action \cite{Hilbert1915}, $q_{ab}$ and $K_{ab}$ result to be the main dynamical variables of a constrained Hamiltonian theory \cite{dirac2001lectures,bojowald_2010}: the metric $q_{ab}$ is configuration variable , while the extrinsic curvature is linked with the conjugate momentum $P^{ab}=\frac{\sqrt{q}}{k}(K^{ab}-q^{ab}K^c_c)$. The lapse function and the shift vector are also configurational variables, but their dynamics is trivial. They play the role of Lagrangian multipliers of the so-called supermomentum $\boldsymbol{D}$ and superHamiltonian $H$, respectively
\begin{align*}
    & \boldsymbol{D}_i=-2q_{ij}\nabla_kP^{kj},\\
    & H=\frac{k}{\sqrt{q}}\left(q_{ik}q_{jl}-\tfrac{1}{2}q_{ij}q_{kl}\right)P^{ij}P^{kl}-\frac{\sqrt{q}}{k}R.
\end{align*}
Here, $\nabla$ is the Levi-Civita covariant derivative associated with $q$ and $R$ is its scalar curvature. These two quantities constitute the constraints $\boldsymbol{D}_i=0$ and $H=0$. Since the lapse function and the shift vector are not involved in the definition of the Ashtekar variables, we will consider the ADM phase space to consist of pairs \((q_{ab}, P^{ab})\), where \(q_{ab}\) is a 3-metric and \(P^{ab}\) is a symmetric tensor on \(\Sigma\), together with the supermomentum and superHamiltonian constraints.
\\\\
The vierbein formulation constitutes another approach to General Relativity. A vierbein is an object, with a coordinate index $\mu$ and an internal index $\alpha$, $e^{\alpha}_{\mu}$ such that $g_{\mu\nu}=\eta_{\alpha\beta}e^{\alpha}_{\mu}e^{\beta}_{\nu}$. The internal index is covariantly derived using the spin connection $\omega_{\mu}^{\alpha\beta}$ requiring 
\begin{equation}
    \label{constraint}
    D_{\mu}e^{\alpha}_{\nu}=\partial_{\mu}e^{\alpha}_{\nu}-\Gamma^{\sigma}_{\mu\nu}e^{\alpha}_{\sigma}+\omega_{\mu\ \beta}^{\ \alpha}e^{\beta}_{\nu}=0.
\end{equation}
This formulation can be adapted to the ADM splitting in the following form
\begin{equation}
\label{vierred}
    e^{\alpha}_{\ \mu}=\begin{pmatrix}
                N & 0 \\
                N^ae^i_a & e^i_a
                \end{pmatrix}
\end{equation}
where $q_{ab}=\delta_{ij}e^i_ae^j_b$. In this case, the $e^i_a$ (called dreibein) emulates the role of the vierbein in the three-dimensional setting and its inverse $e^a_i$ is called triad. The internal index is $i$ and the coordinate index is $a$. Moreover, a three-dimensional version of (\ref{constraint}) holds. This choice for $e^{\alpha}_{\ \mu}$, known in physics as the time gauge, is not the most general. However, it reproduces a canonical construction in mathematics related to embedded hypersurfaces. In particular, its form can be deduced from the canonical embedding of the orthonormal frame bundle of the hypersurface into that of the entire manifold. A brief discussion of this geometric relationship between the vierbein and the dreibein is provided in \ref{AppTetrads}. The time gauge breaks the local Lorentz symmetry by fixing the temporal component of the vierbein to be normal to $\Sigma_t$ and the spatial components to lie in its tangent bundle. For this reason, its role in the quantum theory has been discussed by many authors (e.g. see \cite{Livine_2003,Cianfrani_2007,Cianfrani_2008,Perez_2013} and references therein).\\
In this frame, starting from the Holst action \cite{Holst1996}, it is possible to introduce a new couple of variables $(A^i_a, E^a_i)$, called Ashtekar-Sen connection and electric field (or densitized triad) \cite{ashtekar86,ashtekar87}. Those are conjugate variables of a constrained Hamiltonian theory, the constraints of the theory are
\begin{align}
    \label{Gauss}
    &G_i=\partial_aE^a_i+\epsilon_{ij}^{\ \ k} A^j_aE^a_k,\\
    \label{Diffeo}
    &V_a=E^b_i F^i_{ab},\\
    \label{Hamilt}
    &S=\left(F^i_{ab}-\tfrac{\beta^2+1}{\beta^2}\epsilon^i_{\ jk}(A^j_a-\Gamma^j_a)(A^k_b-\Gamma^k_b)\right)\frac{\epsilon_{imn}E^{a}_{m} E^{b}_{n}}{\sqrt{|\det(E^a_i)|}},
\end{align}
called Gauss, Vector, and Scalar constraints, respectively. Here, $F^i_{ab}=\partial_aA_b^i-\partial_bA_a^i+\epsilon^i_{\ jk}A^j_aA^k_b$ is the curvature of the Ashtekar-Sen connection, while $\Gamma^i_a$ is a homogeneous rational function of degree zero of $E^a_i$ and its derivatives (cf. Eq.4.2.18 in \cite{thiemann_2007}), and $\beta$ is the Barbero-Immirzi parameter that appears as a free parameter in the Holst action \cite{Immirzi_1997}. It is worth noting that the Ashtekar formulation can be derived from the self-dual action without imposing the time gauge \cite{Barbero_2024}.\\
When the Gauss constraint is satisfied, the phase space is equivalent to the ADM formulation with the supermomentum and superHamiltonian constraints \cite{ashtekar87}. Moreover, there exists a map between these two sets of variables
\begin{align}
    \label{AshtConnection}
    &A^i_a=\Gamma^i_a+\beta K^i_a,\\
    \label{DensTriad}
    &E^a_i=\sqrt{q}e^a_i.
\end{align}
where $K^i_a=K_a^be^i_b$, and $\Gamma^i_a=\frac{1}{2}\epsilon^i_{\ jk}\omega^{jk}_a$, with $\omega^{jk}_a$ being the spin connection of the dreibein $e^i_a$. When $\beta$ is a real parameter, the formulation exhibits an $SU(2)$ redundancy; namely, there exists an $SU(2)$ gauge freedom in the choice of the pair $(E^a_i, A^i_a)$ associated with $(q_{ab}, K_{ab})$. In particular, we can interpret the Ashtekar-Sen connection as an $SU(2)$ gauge field (known as the Ashtekar-Barbero-Immirzi-Sen connection, in alphabetical order) \cite{Barbero1995}. Now, we aim to analyze this aspect and formalize this interpretation via the correspondence with connections on a suitable principal $SU(2)$ bundle over the hypersurface $\Sigma$.

\section{Gauge theory structure}
\label{SecGauge}
In this section, we aim to understand how the $SU(2)$ symmetry emerges at the geometric level. Specifically, we will construct a principal $SU(2)$ bundle associated with the metric $q$ and analyze its properties to implement the Ashtekar-Barbero-Immirzi-Sen connection as a connection (in the sense of a principal bundle) on it. For the remainder of the paper, we will refer to the Ashtekar-Barbero-Immirzi-Sen connection simply as the Ashtekar connection, as is common in the literature.\\
Let $(\Sigma,q)$ be a 3-dimensional oriented Riemannian manifold. On such a manifold,  the oriented orthonormal frame bundle $\OF$ is well defined. Consider the fiber on a point $x\in \Sigma$
\[P_x^{SO}(\Sigma)=\{h:\R^3\to \T_x\Sigma\,|\,h\ \textit{orientation preserving linear isometry}\}.\]
The orthonormal frame bundle $\OF\doteq\bigsqcup_{x\in \Sigma}P_x^{SO}(\Sigma)$ is a principal $SO(3)$-bundle over $\Sigma$ with the right action of $SO(3)$ defined by
\begin{align*}
    P_x^{SO}(\Sigma)\times SO(3)\to & P_x^{SO}(\Sigma)\\
    (h, O)\mapsto &\, h\circ O
\end{align*}
Using the construction of the associated vector bundle we can properly see the role of the dreibein.\\\\
The tangent bundle can be realized as an associated vector bundle $\T\Sigma\cong \OF\times_{\lambda_{st}}\R^3$, where $\lambda_{st}:SO(3)\to GL(3,\R)$ is the defining representation of $SO(3)$. Hence, elements of the tangent bundle are equivalence classes of pairs $[h,v],\, h\in\OF,\,v\in \R^3$ associated with the tangent vector $h(v)$. In this context, the index $i$ serves purely as a counting index for the elements of the orthonormal basis. Therefore, it can be raised or lowered using the Kronecker delta.\\
In this setting, the dreibein is equivalent to the data of a (global) section $h:\Sigma\to\OF$. Indeed, given the canonical basis $\{\mathfrak{e}_1,\mathfrak{e}_2,\mathfrak{e}_3\}$ of $\R^3$, we can associate a frame of orthonormal vector fields (the triad): $[h,\mathfrak{e}_i]=h(\mathfrak{e}_i)=:e_i=e^a_i\partial_a$. From this, the cotriad (or dreibein) is the 1-forms dual frame $\{e^i=e^i_adx^a\}\subset \Omega^1(\Sigma)$. Notice that the vector $v$ represents the components of $h(v)$ in the trivialization of the bundle induced by $h$, namely the components with respect to the frame $e_i$.\\
The dreibein is also related to a section in $\mathrm{End}(\T\Sigma)\cong\T^*\Sigma\tensor\T\Sigma$. Actually, the section that to each point associates the Casimir operator of the endomorphisms of the tangent space, namely the identity, can be written as $\mathrm{Id}_{\T\Sigma}=e^a_i e^i\tensor \partial_a=e^i_a dx^a\tensor e_i$. In this context, $e^a_i$ and its inverse $e^i_a$ play the role of change-of-basis matrices. Further details on this aspect can be found in \ref{AppTetrads}.\\

The $SU(2)$ structure emerges when spinors are introduced into the theory. To achieve this, we need to introduce spin structures. The existence of a spin structure on $\Sigma$ can be inferred from the natural requirement of a spin structure on $\M$ (cf. \ref{AppTetrads}). We are going to consider the definition of spin structure given in \cite{Bar}:
\begin{definition}
    \label{spindef}
    A spin structure on $\Sigma$ is a couple $(\SP,\Bar{\rho})$ consisting of:
    \begin{itemize}
        \item a principal $\mathrm{Spin(3)}$-bundle $\SP$ over $\Sigma$,
        \item a twofold covering $\Bar{\rho}:\SP\to \OF$ such that the following diagram commutes:
    \end{itemize}
    \[
    \begin{tikzcd}
    \SP\times \mathrm{Spin}(3) \arrow[r] \arrow[dd,"\Bar{\rho}\times\rho"] & \SP \arrow[dd,"\Bar{\rho}"]\arrow[rd] & \\
       &   & \Sigma\\
    \OF\times SO(3) \arrow[r] & \OF \arrow[ru]& 
    \end{tikzcd}
\]
Here, $\rho:\mathrm{Spin}(3)\to SO(3)$ is the twofold covering homomorphism of $SO(3)$. The horizontal maps are the group operations (right multiplications) on the principal bundles.
\end{definition}

With a spin structure, every associated vector bundle to $\OF$ is isomorphic to one to $\SP$. Consider a representation $\Bar{\lambda}:\mathrm{Spin}(3)\to \mathrm{GL}(V)$ of the form $\Bar{\lambda}=\lambda \circ \rho$, for some representation $\lambda:SO(3)\to \mathrm{GL}(V)$, then $\SP\times_{\Bar{\lambda}}V\cong\OF\times_{\lambda}V$. In particular $\T\Sigma\cong \OF\times_{\lambda_{st}}\R^3\cong \SP\times_{\lambda_{st}\circ \rho}\R^3$. Moreover, in dimension three, since the defining representation of $\mathrm{Spin}(3)$ is equivalent to the adjoint one, we get $\T\Sigma\cong \mathrm{ad}\SP\doteq\SP\times_{\Ad}\mathfrak{spin}(3)$. Thus, the dreibein defines (not uniquely) also a section $\Bar{h}:\Sigma\to \SP$ that satisfies $\Bar{\rho}\circ \Bar{h}=h$. In this way, it defines the same vector fields of $h:\Sigma\to \OF$, in the sense that $e_i=[\Bar{h},\mathfrak{e_i}]\in\SP\times_{\lambda_{st}\circ \rho}\R^3$.\\\\
Moreover, spinnable 3-manifolds are parallelizable. Since $\mathrm{Spin(3)}$ is simply connected and compact, then $\SP$ is a trivial principal bundle over $\Sigma$. As associated bundles of a trivial principal bundle are trivial, $\T\Sigma$ is trivial. Hence, $\OF$ is trivial too. Thus, we can always consider global sections of such bundles.\\\\
From now on, the manifold $\Sigma$ is always considered equipped with a spin structure $(\SP,\Bar{\rho})$ and, to make the discussion closer to the usual physical notation, we identify $\mathrm{Spin}(3)\cong SU(2)$, and so $\mathfrak{spin}(3)\cong \su{2}$.

\subsection{Connection and gauge field}
\label{SubConnection}
Connections are the central objects in gauge theories. In this paper, we consider connections as 1-forms on the principal bundle, taking values in the Lie algebra of the structure group \cite{Kobayashi1}.\\\\
The connections on $\OF$ are the so-called metric-compatible connections. Among them, we can recognize a preferred one which is the Levi-Civita connection $\omega^{LC}$. The spin structure allows us to define a one-to-one correspondence between connections on $\OF$ and connections on $\SP$. Given a connection $\omega$ on $\OF$, we can define a connection $\Bar{\omega}=\rho^{-1}_*\Bar{\rho}^*\omega$ on $\SP$, where $\rho^{-1}_*$ is the isomorphism $\so{3}\xrightarrow{\sim}\su{2}$ induced by $\rho$. The contrary is also true, every connection on $\SP$ can be written in this way.\\\\
The Ashtekar connection is the local 1-form of a connection on $\SP$, namely its pullback via a section
\begin{equation}
    A=\Bar{h}^*\Bar{\omega}.
\end{equation}
The Ashtekar connection $A$ is a $\su{2}$-valued 1-form $A=A^i_a dx^a\tensor \tau_i$, where $\tau_i$ is a basis of generators for $\su{2}$. Under a change of section $\Bar{h}\mapsto \Bar{h}g$, with $g:\Sigma\to SU(2)$, transforms $A\mapsto \Ad_{g^{-1}}A +g^{-1}dg$. It is clear that, due to the properties of the spin structure, to deal with an Ashtekar connection is equivalent to treating with a connection in $\OF$. Actually, since the bundle maps $\bar h^*$ and $\bar\rho^*$ commute with the Lie algebra isomorphism $\rho_*$, $A$ is also the local 1-form of a metric-compatible connection \[A=\Bar{h}^*\Bar{\omega}=\Bar{h}^*\rho^{-1}_*\Bar{\rho}^*\omega=\rho^{-1}_*(\Bar{\rho}\circ \Bar{h})^*\omega=\rho^{-1}_*h^*\omega.\]
In which the only difference is that $\tau_i$ is substituted by $T_i$, the generators of $\so{3}$. In such a way, the local field $A$ depends only on the section $h$ and not on the specific lift $\Bar{h}$, and in general by the chosen spin structure.\\\\
A curvature form $\Theta$ is associated with a connection $\Bar{\omega}$. It is a 2-form on $\SP$ with values in $\su{2}$ defined by: \[\Theta=d\Bar{\omega}+\frac{1}{2}\ad(\Bar{\omega})\wedge\Bar{\omega}.\]
The pullback of $\Theta$ is the curvature (or field strength) $F=\Bar{h}^*\Theta$. The curvature is written locally as
\begin{equation}
    F=dA+\tfrac{1}{2}\ad(A)\wedge A.
\end{equation}
The curvature transforms under a change of section $\Bar{h}\mapsto \Bar{h}g$ as $F\mapsto \Ad_{g^{-1}}F$ and in a coordinate chart reads:
\begin{align*}
    F&=(\partial_aA^i_b \tau_i + \tfrac{1}{2}A^j_aA^k_b[\tau_j,\tau_k])dx^a \wedge dx^b=(\partial_aA_b^i+\tfrac{1}{2}\epsilon^i_{\ jk}A^j_aA^k_b)\tau_i\tensor dx^a \wedge dx^b\\
    &=\tfrac{1}{2}(\partial_aA_b^i-\partial_bA_a^i+\epsilon^i_{\ jk}A^j_aA^k_b)\tau_i\tensor dx^a \tensor dx^b=\tfrac{1}{2}F^i_{ab}\tau_i\tensor dx^a \tensor dx^b.
\end{align*}
We can associate with $\Theta$ an element of $\Omega^2(\Sigma,\mathrm{ad}\SP)$ uniquely. Explicitly, it is $\boldsymbol{F}=[\Bar{h},\tfrac{1}{2}F^i_{ab}\tau_i]\tensor dx^a \wedge dx^b$. Hence, $F$ is the expression of $\boldsymbol{F}$ in the trivialization of the bundle $\ad\SP$ induced by the section $\Bar{h}$. For simplicity, by now we will omit the $k$-form part in this kind of object, so the previous reads $\boldsymbol{F}=[\Bar{h},F^i\tau_i]=[\Bar{h},F]$.

\subsection{Electric field and solder form}
\label{SubEle}
 In physics, the dreibein is collected in a $\su{2}$-valued 1-form. Given the cotriad $e^i$, we are going to consider the contraction of the 1-forms with the basis of generators $\tau_i$ of $\su{2}$ \[e=e^i_adx^a\tensor \tau_i.\]
 This construction, apparently unnatural, is linked with a solder form on $\SP$.\\
 
 On the orthonormal frame bundle, the tautological solder form $\theta$ is a $1$-form on $P^{SO}(\Sigma)$ with values in $\R^3$ defined by
\[\theta_{(x,h)}(v)=h^{-1}\left(d_{(x,h)}\pi (v)\right),\]
where $x\in\Sigma,h\in P^{SO}_{x}(\Sigma), v\in \T_{(x,h)}P^{SO}(\Sigma)$ and $\pi$ is the projection of the bundle. The solder form is right equivariant $R_g^*\theta=\lambda_{st}(g^{-1})\theta$. Let $h$ be a section of $P^{SO}(M)$, then the cotriad is 
\begin{equation}
    (e^1,e^2,e^3)=h^*\theta.
\end{equation}
Thus, point by point, for all $v\in \T_xM$
\begin{align*}
    e^i_a(x)dx^a(v)=h_x^{-1}\left(d_{(x,h_x)}\pi \circ d_xh(v)\right)=h_x^{-1}\left(d_x(\pi \circ h)(v)\right)=h_x^{-1}(v)
\end{align*} 

From the tautological solder form, we can induce a solder form on $\SP$. Considering firstly the pullback of $\theta$ via $\Bar{\rho}$ and then the composition with the $SU(2)$-equivariant isomorphism $\phi:\R^3\xrightarrow{\sim}\su{2}$, the solder form $\vartheta\in\Omega^1(\SP,\su{2})$ is defined by:
\begin{equation}
    \vartheta\doteq \phi\circ \Bar{\rho}^*\theta.
\end{equation}
 It is horizontal because $\theta$ is horizontal. Moreover, it is $SU(2)$-equivariant:
 \begin{align*}
     R_g^*\vartheta=\phi\circ R_g^*\Bar{\rho}^*\theta=\phi\circ \Bar{\rho}^*R_{\rho(g)}^*\theta=\phi (\lambda_{st}(\rho(g^{-1}))\Bar{\rho}^*\theta)=\Ad_{g^{-1}}\phi (\Bar{\rho}^*\theta).
 \end{align*}
 It is well-defined since the adjoint action of $SU(2)$ is equivalent to the adjoint action of $SO(3)$. With such a form, it is possible to associate uniquely a differential 1-form $\boldsymbol{e}\in\Omega^1(\Sigma, \mathrm{ad}\SP)$ that, in a trivialization $\Bar{h}$, reads as
 \begin{equation}
     e=\Bar{h}^*\vartheta.
 \end{equation}
 From which, the transformation under change of trivialization is $e\mapsto \Ad_{g^{-1}}e$.\\\\
 Now, we can introduce the electric field $E$. In a trivialization of $\SP$, it can be defined as an element of $\Omega^2(\Sigma,\su{2}^*)$ such that
 \begin{equation}
 \label{defE}
     \frac{1}{3}tr(E\wedge e)=\dV,
 \end{equation}
 where $tr$ is the natural pairing between $\su{2}^*$ and $\su{2}$ (or, analogously, the multiple of Killing form such that the generators of $\su{2}$ form an orthonormal basis. In particular, it is proportional to the standard trace $\mathrm{Tr}$, $tr=-2\mathrm{Tr}$). It is clear that $E$ is the contraction of the Hodge dual of the cotriad
\begin{equation}
\label{HodgeE}
    \star e^i=\tfrac{1}{2}\sqrt{q}e^i_{a'}q^{a'a}\epsilon_{abc}dx^b\wedge dx^c=\tfrac{1}{2}\delta^{ij}E^a_j\epsilon_{abc}dx^b\wedge dx^c,
\end{equation}
with the dual basis of $\su{2}$ given by $\{\mathfrak t^i\}\subset\su{2}^*$. Under a change of trivialization $\Bar{h}\mapsto\Bar{h}g$, the electric field must transform as $E\mapsto \Ad_{g^{-1}}^*E$, where $\Ad^*$ is the coadjoint representation, to maintain the invariance of Eq.(\ref{defE}).\\
Hence, $[\Bar{h},\tfrac{1}{2}E^a_i\epsilon_{abc}\mathfrak{t}^i]\tensor dx^b\wedge dx^c$ is a well-defined 2-form with values in $\mathrm{ad}^*\SP$, namely $\boldsymbol{E}=[\Bar{h},E]\in\Omega^2(\Sigma,\mathrm{ad}^*\SP)$. Such a vector field is associated with a horizontal $2$-form $\tau\in\Omega^2(\SP,\su{2}^*)$ such that $\boldsymbol{E}=[\Bar{h},\Bar{h}^*\tau]$.

\subsection{Equivalence of the data}
\label{SubRecostruction}

We want to show how to reconstruct a $(0,2)$-tensor field $\mathrm{K}$ starting from a connection $\omega$ on $\SP$.\\
On $\SP$, there exists a special connection that is the spin connection $\Bar{\omega}^{LC}$ associated with the Levi-Civita connection $\Bar{\omega}^{LC}=\rho^{-1}_*\Bar{\rho}^*\omega^{LC}$. The difference $\omega-\Bar{\omega}^{LC}$ is a horizontal and $SU(2)$-invariant 1-form on $\SP$, so, it is associated with an element of $\Omega^1(\Sigma,\mathrm{ad}\SP)$. In a trivialization $\Bar{h}$, it reads $\beta K=\Bar{h}^*(\omega-\Bar{\omega}^{LC})$, where $\beta\in\R$. From the global trivialization, we can define the tensor
\begin{equation}
    \mathrm{K}=tr(e\tensor K)
\end{equation}
where $e=\Bar{h}^*\vartheta$ is the solder form. $\mathrm{K}$ is invariant under change of trivialization since $tr$ is proportional, pointwise, to the Killing form on $\su{2}$, then $tr(\Ad_g e\tensor \Ad_g K)=tr(e\tensor K)$ for all $g:\Sigma\to SU(2)$. Therefore, $\mathrm{K}\in\Omega^1(\Sigma)\tensor\Omega^1(\Sigma)$ is a well-defined tensor.\\
For an \textit{a priori} trivialization-free definition, we must use the bundle isomorphism $\Phi:\T\Sigma\xrightarrow{\sim}\mathrm{ad}\SP $ induced by the solder form $\vartheta$ and defined by $\Phi: X\mapsto \boldsymbol{e}(X)$. Then, we can define $\mathrm{W}=\Phi^{-1}([\Bar{h},\beta\Bar{h}^*(\Bar{\omega}^{A}-\Bar{\omega}^{LC})])$ and conclude $\mathrm{K}(X,Y)=q(\mathrm{W}(X),Y)\,\forall X,Y\in \T_x\Sigma$.\\
Finally, we must check that, in local coordinates, $\mathrm{K}$ is the map (\ref{AshtConnection}): 
\begin{align*}
    \beta\mathrm{K}&=tr(e\tensor \beta K)=e^j\tensor(A^i-\Gamma^i) tr(\tau_i\tau_j)=e^j\tensor(A^i-\Gamma^i) \delta_{ij}\\
    &=e^i_a(A^j_b-\Gamma^j_b)\delta_{ij}dx^a\tensor dx^b.
\end{align*}
However, the symmetry of this tensor is not deduced from this construction. The Gauss constraint ensures this property.

\section{Constraints}
\label{SecConst}
The constraints arise from the Hamiltonian formulation of General Relativity in Ashtekar variables, and they have a well-known interpretation as the generators of infinitesimal transformations of the phase space \cite{thiemann_2007}. However, they can also acquire a geometric interpretation and can be described within the framework of fiber bundle theory.\\\\
We recall that associated with a connection on a principal bundle there exists a unique covariant derivative on the associated vector bundles. Hence, a connection $\omega$ on $\SP$ induces a covariant derivative $D$ on $\T\Sigma$ and its functorial constructions \cite{Demailly_1997}.  Its action on the triad and its dual frame reads in terms of Ashtekar connection
\begin{align*}
    &De_i=[\Bar{h},d\mathfrak{e}_i+\lambda_*(\Bar{h}^*\omega)\mathfrak{e}_i]=[\Bar{h},A^j\lambda_*(\tau_j)\mathfrak{e}_i]=[\Bar{h},A^j\epsilon_{jik}\mathfrak{e}_k]=A^j\epsilon_{ji}^{\ \ k}e_k\\
    &De^i=[\Bar{h},d\mathfrak{e}_i-\prescript{t}{}{\lambda}_*(\Bar{h}^*\omega)\mathfrak{e}_i]=[\Bar{h},-A^j\prescript{t}{}{\lambda}_*(\tau_j)\mathfrak{e}_i]=[\Bar{h},A^j\epsilon_{jik}\mathfrak{e}_k]=A^j\epsilon_{jik}e^k.
\end{align*}
Here, $-\prescript{t}{}{\lambda}_*$ is the dual representation, that, in terms of matrices is minus the transpose of the defining representation $\lambda_*$ of the Lie algebra $\su{2}$. The indices $i,j,k$ are raised and lowered using the Kronecker delta. Moreover, let $\nabla$ be the  Levi-Civita covariant derivative, we recall that $\nabla e_i=\Gamma^k\epsilon_{ki}^{\ \ j} e_j.$\\
A connection $\omega$ induces also the so-called exterior covariant derivative $d_{\omega}$ on $k$-forms with values in a vector bundle. The derivation rule is the following: consider $\eta$ a $k$-form on $\Sigma$ and $s$ a section in a associated vector bundle
\[d_{\omega}(\eta\tensor s)=d\eta\tensor s+(-1)^k\eta\wedge Ds.\]
Notice that the $\su{2}$-valued 1-form $K$, defined in Sec.\ref{SubRecostruction}, plays a role in the induced covariant derivative. Indeed, the covariant derivative on $\T\Sigma$ can be written as $D=\nabla+\mathbf{K}$, where $\mathbf{K}\in \Omega^1(\Sigma,\mathrm{AntiSym}(\T\Sigma))$. Here, $\mathrm{AntiSym}(\T\Sigma)$ is the bundle of skew-symmetric linear operators on the tangent space. In particular, its action on the dreibein is
\begin{align*}
    \mathbf{K}e_i&=De_i-\nabla e_i=[\Bar{h},d\mathfrak{e}_i+\lambda_*(\Bar{h}^*\omega)\mathfrak{e}_i]-[\Bar{h},d\mathfrak{e}_i+\lambda_*(\Bar{h}^*\Bar{\omega}^{LC})\mathfrak{e}_i]\\
    &=[\Bar{h},\lambda_*(\Bar{h}^*(\omega-\Bar{\omega}^{LC}))\mathfrak{e}_i]=[\Bar{h},\lambda_*(\beta K)\mathfrak{e}_i]=[\Bar{h},\beta K^j\epsilon_{jik}\mathfrak{e}_k]=\beta K^j\epsilon_{ji}^{\ \ k}e_k.
\end{align*}
From which $q(\mathbf{K}(\partial_a)e_i,e_j)=\beta K_a^l\epsilon_{lij}$, showing the skew-symmetry property.

\subsection{Gauss constraint}
\label{SubGauss}
It is well-known that the Gauss constraint forces the tensor $\mathrm{K}$ to be symmetric \cite{thiemann_2007}. This fact follows from a simple computation, for completeness we reproduce it here
\begin{align*}
    G_i&=\partial_aE^a_i+\epsilon_{ij}^{\ \ k}A^j_aE^a_k=\sqrt{q}\left(\nabla_a e^a_i+\epsilon_{ij}^{\ \ k}A^j_ae^a_k \right)=\sqrt{q}\left(\Gamma_a^j\epsilon_{ji}^{\ \ k} e^a_k+\epsilon_{ij}^{\ \ k}A^j_ae^a_k \right)\\
    &=\sqrt{q}\left(-\beta K_a^j\epsilon_{ji}^{\ \ k} e^a_k\right)=\beta\sqrt{q} K_{ab}\epsilon_{i}^{\ jk} e^a_ke^b_j
\end{align*}
It is quite evident that \[G_i=0\ \iff\ \epsilon^{ikj}K_{ab} e^a_ke^b_j=0\ \iff\ K_{ab}=K_{ba}.\]
We are now interested in a coordinate-free definition and a geometric interpretation of the Gauss constraint. It can be found quite straightforward via the exterior covariant derivative of the electric field:
\begin{equation}
    \boldsymbol{G}=d_{\omega}\boldsymbol{E}.
\end{equation}
This defines an element of $\Omega^3(\Sigma,\mathrm{ad}^*\SP)$. Considering a global trivialization, we can define the $\su{2}^*$-valued 3-form
\begin{equation}
    G=d_AE\doteq dE+\ad^*(A)\wedge E.
\end{equation}
In such a way, $\boldsymbol{G}=[\Bar{h},G]$. The proof that this object is the usual Gauss constraint follows immediately:
\begin{align*}
    &d_{\omega}\boldsymbol{E}=d(\tfrac{1}{2}E^a_i\epsilon_{abc}dx^b\wedge dx^c)\tensor[\Bar{h},\mathfrak t^i]+(\tfrac{1}{2}E^a_i\epsilon_{abc}dx^b\wedge dx^c)\wedge D[\Bar{h},\mathfrak t^i]\\
    &=(\tfrac{1}{2}\partial_{a'}E^a_i\epsilon_{abc}dx^{a'}\wedge dx^b\wedge dx^c)\tensor[\Bar{h},\mathfrak t^i]+\tfrac{1}{2}E^a_i\epsilon_{abc}dx^b\wedge dx^c\wedge dx^{a'}\tensor[\Bar{h},A^j_{a'}\epsilon_{jik}\mathfrak t^k]\\
    &=(\partial_aE^a_i+\epsilon_{ijk}A^j_aE^a_k)dx^1\wedge dx^2 \wedge dx^3\tensor[\Bar{h},\mathfrak t^i]=G_idx^1\wedge dx^2 \wedge dx^3\tensor[\Bar{h},\mathfrak t^i].
\end{align*}
Thus, the constraint $\boldsymbol{G}=0$ imposes the vanishing of the exterior covariant derivative for the electric field, and the equation $d_{A}E=0$ is analogous to the equations of motion of a Yang-Mills theory.

\subsection{Vector constraint}
\label{SubDiffeo}
The Vector constraint admits an expression in terms of exterior covariant derivative too. Its expression in terms of the cotriad is clear and concise and gives us an element of $\Omega^3(\Sigma,\mathrm{ad}\SP)$
\begin{equation}
\label{DiffC}
    \boldsymbol{V}=d_{\omega}^2\boldsymbol{e}.
\end{equation}
In a global trivialization, its components read as
\begin{equation}
    \mathcal{V}=d^2_Ae=\ad(F)\wedge e.
\end{equation}
However, in physics, the Vector constraint is a 1-form. We can obtain this formulation considering the bundle isomorphism induced by the solder form, analogously to Sec.~\ref{SubRecostruction}. In this case, we can define $\mathrm{V}\in\Omega^3(\Sigma)\tensor\Omega^1(\Sigma)$ as
\begin{equation}
    \mathrm{V}=tr(e\tensor \mathcal{V}).
\end{equation}
Now, we can check that this expression of the Vector constraint is the usual one by performing a computation in coordinates
\begin{align*}
    \mathcal{V}=\ad(F)\wedge e=F^i\wedge e^j\tensor[\tau_i,\tau_j]=\tfrac{1}{2}\epsilon_{ijk}F^i_{ab}e^j_c dx^a\wedge dx^b\wedge dx^c \tensor\tau_k,
\end{align*}
and then
\begin{align*}
    \mathrm{V}&=e^l\tensor\epsilon_{ijk}F^i\wedge e^j tr(\tau_l\tau_k)=\tfrac{1}{2}\epsilon^{abc}\epsilon_{ijk}F^i_{ab}e^j_ce^k_{a'} dx^1\wedge dx^2\wedge dx^2 \tensor dx^{a'}\\
    &=\tfrac{1}{2}\epsilon^{abc}F^i_{ab}\epsilon_{a'b'c}E^{b'}_i dx^1\wedge dx^2\wedge dx^2 \tensor dx^{a'}=F^i_{ab}E^{b}_i dx^1\wedge dx^2\wedge dx^2 \tensor dx^{a}.
\end{align*}
Where we used the property $E^i_a=\frac{1}{2}\epsilon^{ijk}\epsilon_{abc}e^b_je^c_k$, obtained by the Hodge duality (\ref{HodgeE}). Clearly, $\mathcal{V}=0$ if and only if $\mathrm{V}=0$.\\
The equation $d^2_A e=0$ does not resemble one from Yang-Mills theory, even considering $e$ the ``Hodge dual" of $E$ (see Eq.(\ref{defE})). In this context, a Yang-Mills-like equation is $d_Ae=0$, and if this equation holds, then the Vector constraint is necessarily satisfied. This Yang-Mills-like equation is the local version of the torsion-less condition for the connection $\omega$ because $e$ is the local realization of the solder form $\vartheta$. Hence, it can be written as a pullback, with respect to a global section, of the equation on $\SP$ 
\begin{equation}
    d\vartheta+\ad(\omega)\wedge\vartheta=0.
\end{equation}
The left-hand side of this equation is the torsion form $\mathfrak T$ of $\omega$. Thus, the Vector constraint requires a connection $\omega$ with parallel torsion form, namely a torsion form with vanishing exterior covariant derivative:
\begin{equation}
    d_{\omega}\mathfrak{T}=\ad(\Theta)\wedge\vartheta=0.
\end{equation}

Now, we show that the Vector constraint yields the Codazzi equation when the Gauss constraint is satisfied. This is a standard result that we are going to prove in this bundle language.\\ 
First of all, we can rewrite the curvature $F$ in a fixed trivialization splitting $\Gamma$ and $K$:
\begin{equation}
\label{Fdec}
    F=R+\beta dK+\beta \ad(\Gamma) \wedge K+\tfrac{\beta^2}{2}\ad(K)\wedge K.
\end{equation}
Here, $R$ is the curvature of the Levi-Civita connection $\Gamma$. Therefore, the Vector constraint is composed by four terms. The first term $\ad(R)\wedge e$ vanishes because it is the Vector constraint for the connection $\Gamma$, but it is the Levi-Civita connection and so it is torsion-free.\\
The fourth term $\ad(\ad(K)\wedge K)\wedge e=2\ad(K)\wedge\ad(K)\wedge e$ is proportional to the Gauss constraint. Thus, considering the Gauss constraint satisfied, the only non-trivial contribution to the Vector constraint comes from the middle terms of $F$, which are the exterior covariant derivative of $K$ with respect to the Levi-Civita connection. Hence, the reduced Vector constraint reads
\begin{equation}
    \ad(d_{\Gamma}K)\wedge e=0.
\end{equation}
Passing to a coordinate system, we obtain the following equation 
\[\left(\partial_aK_b^i-\partial_bK_a^i+\epsilon^i_{\ jk}\Gamma^j_aK^k_b+\epsilon^i_{\ jk}K^j_a\Gamma^k_b\right)E^b_i=0.\]
Using the relation with the usual Christoffel symbols $\Gamma^k_a\epsilon_{kj}^{\ \ i}={e^i_c}\Gamma^c_{ab}e^b_j+e^i_b\partial_ae^b_j$ the previous equation reads
\begin{equation}
    \sqrt{q}\left(\partial_aK_b^b-\partial_bK_a^b+\Gamma^b_{ac}K^c_b-\Gamma^c_{bc}K^b_a\right)=0.
\end{equation}
This equation is known as the contracted Codazzi equation \cite{Kobayashi2} in a Ricci flat spacetime.

\subsection{Scalar constraint}
\label{SubScal}
Even for the Scalar constraint, we can give a formulation in terms of connection and electric field. The Scalar constraint can be written as
\begin{equation}
\label{ScalC}
S=tr\left( E\wedge\star\big(F-\tfrac{1}{2}(\beta^2+1)\ad(K)\wedge K\big)\right).
\end{equation}
Hence, $S$ is a top-form, namely an element of $\Omega^3(\Sigma)$. Here, $K$ is a function of $A$ and the dreibein as defined in Sec.\ref{SubRecostruction}. The Scalar constraint is interpreted as a section of the trivial bundle \(\Sigma \times \mathbb{R}\), which is an associated bundle of $\SP$ with the vector space \(\R\) under the trivial representation \(\lambda_{\text{tr}}: SU(2) \to GL(\mathbb{R})\), mapping every group element to the identity element on $\R$. Since sections of \(\Sigma \times \mathbb{R}\) are canonically identified with real-valued functions on \(\Sigma\), we have 
\[
\Omega^3(\Sigma, \Sigma \times \R) \cong \Omega^3(\Sigma) \otimes \Omega^0(\Sigma) \cong \Omega^3(\Sigma).
\]\\
The computation that Eq.(\ref{ScalC}) yields Eq.(\ref{Hamilt}) in a local coordinate system is quite simple, and we will show it only for the so-called Euclidean part:
\begin{align*}
tr(E\wedge\star F)&=\tfrac{1}{4}\epsilon_{ijk}F^i_{ab}e^a_le^b_m\epsilon^{lm}_{\ \ n}e^j\wedge e^k\wedge e^n=\tfrac{1}{2}\epsilon_{ijk}\epsilon^{jkn}F^i_{ab}e^a_le^b_m\epsilon^{lm}_{\ \ n}\dV\\
&=\tfrac{1}{2}\epsilon^{lm}_{\ \ i}F^i_{ab}e^a_le^b_m\dV=\frac{1}{2}\epsilon_i^{\ jk}F^i_{ab}\frac{E^a_jE^b_k}{\sqrt{q}}dx^1\wedge dx^2\wedge dx^3.
\end{align*}
We can also notice that the definition of scalar curvature from Riemannian geometry provides a similar result
\begin{align*}
\mathrm{scal}_F\doteq&\sum_{i,j}q(D^2_{(e_i,e_j)}e_i,e_j)=\sum_{i,j,k,l}\frac{1}{2}F_{ab}^ke^a_ie^b_j \epsilon_{kil}q(e_l,e_j)\\
=&\sum_{i,j,k}\frac{1}{2}F_{ab}^k\epsilon_{kij}e^a_ie^b_j.
\end{align*}
Thus, $tr(E\wedge\star F)=\mathrm{scal}_F\dV$.\\\\
From the Scalar constraint, we can extract the Gauss equation using the decomposition (\ref{Fdec}) and imposing the Gauss constraint (i.e. considering a symmetric $\mathrm{K}$). The $\beta^2$ terms cancel out while the other terms must be studied one by one. The first term $tr(E\wedge\star R)=\mathrm{scal}_R\dV$ is the Ricci curvature. The second term $tr(E\wedge \star d_{\Gamma}K)=tr(e\wedge d_{\Gamma}K)=d\,tr(e\wedge K)+tr(d_{\Gamma}e\wedge K)$ vanishes because $tr(e\wedge K)=0$ when $\mathrm{K}$ is symmetric and $d_{\Gamma}e=0$ since $\Gamma$ is torsion-less. The last term is
$tr(E\wedge \star(\ad(K)\wedge K))$ that is the same of $(\sum_{i,j}\mathrm{K}(e_i,e_j)^2-\sum_i\mathrm{K}(e_i,e_i)^2)\dV$. Thus, the constraint $S=0$ reads
\begin{equation}
    \mathrm{scal}_R+\sum_{i,j}\mathrm{K}(e_i,e_j)^2-\sum_i\mathrm{K}(e_i,e_i)^2=0
\end{equation}
This equation is the Gauss equation for the scalar curvature in a Ricci flat spacetime \cite{Kobayashi2}. 

\section{Elements from the physical theory}
A crucial property of General Relativity is background independence. In our description, we fixed a metric in order to construct the spin structure and the related geometrical quantities. However, we want now to describe the metric as a dynamical quantity reconstructed from the electric field. Indeed, the relevant geometric quantities are objects on $\SP$ independent from the spin structure and from the metric, such as the physical variables. Therefore, we now aim to emancipate from the metric by encoding the metric data, and even the orientation, in the electric field, using the correspondence between $q$ and $\boldsymbol{E}$, or equivalently $\tau$ (cf. Sec. \ref{SubEle}), and prove the independence of the set of connections from the metric.\\
Thus, on a spinnable closed 3-manifold $\Sigma$, our data will consist of the pair $(\omega, \tau)$, which allows us to reconstruct the pair of tensors $(q, K)$. Here, $\omega$ is a connection on the unique principal $SU(2)$ bundle $\SP$ over $\Sigma$, while $\tau$ is a non-degenerate element of $\Omega^2(\SP) \otimes \su{2}^*$ that is horizontal and $SU(2)$-equivariant.

\subsection{Independence from the metric}
\label{SubInd}
In the previous Sections, we used the spin structure as a necessary tool to give a geometrical interpretation to the physical variables and to prove the equivalence of the data. To do that, we fixed a metric. However, we will prove that the set of Ashtekar connections is actually independent from the metric.\\
Fixing a spin structure on a given metric, fixes the spin structure for every other metrics. Let us consider a metric $q$ associated with an orthonormal frame bundle $P^{SO}(\Sigma,q)$ and with a spin structure $(\SP,\Bar{\rho})$. A different metric $q'$ on $\Sigma$ is associated with a different orthonormal frame bundle $P^{SO}(\Sigma,q')$. Between these two principal bundles, there exists a vertical isomorphism $\psi:P^{SO}(\Sigma,q)\to P^{SO}(\Sigma,q')$, given pointwise by the matrix which maps an orthonormal basis of $q$ in an orthonormal basis of $q'$. Moreover, there exists a vertical automorphism $\Bar{\psi}:\SP\to\SP$. Defining $\Bar{\rho}':\SP\to P^{SO}(\Sigma,q')$ such that the following diagram commutes:
\[
\begin{tikzcd}
    \SP \arrow[rr, "\Bar{\psi}"] \arrow[d, "\Bar{\rho}"] & & \SP \arrow[d, "\Bar{\rho}' " left]\\
    P^{SO}(\Sigma,q) \arrow[rr, "\psi"] \arrow[rd, ] & & P^{SO}(\Sigma,q') \arrow[ld, ]\\
     & \Sigma & 
\end{tikzcd}
\]
The couple $(\SP,\Bar{\rho}'=\psi\circ\Bar{\rho}\circ \Bar{\psi}^{-1})$ clearly defines a spin structure for $P^{SO}(\Sigma,q')$. Furthermore, the isomorphism $\psi$ defines an isomorphism between the sets of metric-compatible connections, namely for any connection $\omega$ on $P^{SO}(\Sigma,q)$ there exists a unique connection $\omega'$ on $P^{SO}(\Sigma,q')$ such that $\omega=\psi^*\omega'$. For any couple of connections of this kind, the respective spin connections are linked by a gauge transformation $\Bar{\rho}^*\omega=\Bar{\psi}^*\Bar{\rho}'^*\omega'$. Thus, the space of spin connections does not depend on the metric.

\subsection{Reconstructing the metric}
\label{SubMetr}
Fixed a global trivialization of $\SP$, the non-degeneracy of $\tau$ means that the three 2-forms $E_i\in\Omega^2(\Sigma)$ that characterize $E=\Bar{h}^*\tau=E_i\mathfrak{t}^i$ are linearly independent. Hence, they define a global frame for $\Lambda^2\T^*\Sigma$. In such a way, it is possible to induce a scalar product and an orientation on $\T\Sigma$.\\

To show that, let us see what happens in $\R^3$. Let us consider a basis $\{W_i\}\subset \Lambda^2\R^3$. Each vector of $\Lambda^2\R^3 $ can be decomposed as $W=u\wedge v$ with $u,v \in \R^3$. We want to prove that the basis is decomposed uniquely on a basis of $\R^3$.\\
There exist $u_1,u_2\in\R^3$ such that $W_1=u_2\wedge u_3$, we can complete $\{u_1,u_2\}$ to a basis for $\R^3$ adding the vector $u_3$, then  $W_2$ is a linear combination $W_2=\alpha u_1\wedge u_2+\beta u_2\wedge u_3+\gamma u_3\wedge u_1$. We have two cases:
\begin{enumerate}
    \item If $\alpha\neq 0$, we can substitute $u_1$ with the vector $u_1+\frac{\beta}{\alpha}u_3$. And then $u_2$ with $u_2+\frac{\gamma}{\alpha}u_3$, obtaining $W_2=\alpha u_1\wedge u_2$ and leaving $W_1$ invariant.
    \item If $\gamma\neq 0$, we can substitute $u_1$ with the vector $u_1+\frac{\beta}{\gamma}u_2$. And then $u_3$ with $u_3+\frac{\alpha}{\gamma}u_2$, obtaining $W_2=\gamma u_3\wedge u_1$ and leaving $W_1$ invariant.
\end{enumerate}
The two results are linked to a choice of orientation for the basis $\{W_1,W_2,W_3\}$, so, without loss of generality, we can consider $\gamma\neq 0$. The last vector is again a linear combination $W_3=\alpha' u_1\wedge u_2+\beta' u_2\wedge u_3+\gamma' u_3\wedge u_1$, where necessarily $\alpha'\neq 0$. We can substitute $u_1$ with the vector $u_1+\frac{\beta}{\alpha}u_3$. And then $u_2$ with $u_2+\frac{\gamma}{\alpha}u_3$, obtaining $W_3=\alpha' u_1\wedge u_2$ and leaving $W_1$ and $W_2$ invariant. The last step is the rescaling of $u_2$ by $\sqrt{\frac{\alpha'}{\gamma}}$ and $u_3$ by $\sqrt{\frac{\gamma}{\alpha'}}$, the radicand is positive under a proper choice of the orientation, i.e. a relabeling of $W_i$. Thus we obtain that, given a basis $\{W_1,W_2,W_3\}\subset\Lambda^2\R^3$ there exists a basis $\{u_1,u_2,u_3\}\subset\R^3$ such that $W_1=u_2\wedge u_3,\,W_2= u_3\wedge u_1,\,W_3= u_1\wedge u_2$.\\
These bases are metric compatible in the sense that if $\{u_i\}$ is an orthonormal basis then $\{W_i\}$ is an orthonormal basis with respect to the scalar product induced by the one on $\R^3$. The proof is straightforward
\begin{align*}
    \bra W_i,W_j\ket&=\frac{1}{4}\epsilon_{ilk}\epsilon_{jmn}\bra u_l\wedge u_k,u_m\wedge u_n\ket=\frac{1}{4}\epsilon_{ilk}\epsilon_{jmn}(\delta_{lm}\delta_{kn}-\delta_{ln}\delta_{km})\\
    &=\frac{1}{2}\epsilon_{ilk}\epsilon_{ilk}=\delta_{ij}
\end{align*}
Moreover, the decomposition is unique. Suppose there exists $u'_i=\alpha_{ik}u_k$ such that $W_i=\frac{1}{2}\epsilon_{ijk}u_j\wedge u_k=\frac{1}{2}\epsilon_{ijk}u'_j\wedge u'_k$, then the coefficients $\alpha_{ij}$ must satisfy \[\det(\alpha)\epsilon_{ijk}=\epsilon_{ljk}\alpha_{li}.\]
Forcing the matrix $\alpha=\{\alpha_{nm}\}$ to be diagonal and $\det(\alpha)$ to be the diagonal terms. Hence, the only solution is the identity matrix.\\
Moreover, if two bases $\{W_i'\},\{W_i\}\subset\Lambda^2\R^3$ are linked by an orthogonal transformation, the two bases of the decomposition are linked by the same transformation.\\ 

Hence, since the fiber of \(\T^*\Sigma\) is isomorphic to \(\R^3\), for each \(E \in \Omega^2(\Sigma) \otimes \mathfrak{su}(2)\), there exists a unique coframe of 1-forms \(\{e^i\} \subset \Omega^1(\Sigma)\) that decomposes the frame of 2-forms \(\{E_i\} \subset \Omega^2(\Sigma)\) as $E_i=\frac{1}{2}\epsilon_{ijk}e^j\wedge e^k$. The 1-forms frame $ e^i$ is the reconstructed dreibein or cotriad. Its dual frame \(\{e_i\} \subset \mathfrak{X}(\Sigma)\) is the reconstructed triad and, declaring it to be orthonormal, is associated with a unique metric \(q\). Moreover, it induces an orientation on $\T\Sigma$.\\
This procedure, starting from \(\tau\), defines a unique metric \(q\) and provides an orientation for $\Sigma$. Changing the global trivialization, \(E\) changes via the adjoint representation, and thus the frame \(\{E_i\}\) is rotated by a special orthogonal matrix. As a result, the dreibein is also rotated by the same special orthogonal matrix, meaning that the associated metric and the orientation do not change. Even considering a different set of generators of \(\mathfrak{su}(2)\), the metric remains invariant. Indeed, a change of generators corresponds to an automorphism of the Lie algebra, but \(\mathrm{Aut}(\mathfrak{su}(2)) = SO(3)\), leading to the same conclusion.

\subsection{Symmetries}
An important aspect of the physical theory is its symmetries. We now aim to analyze the known symmetries of the Ashtekar-Barbero-Immirzi formulation of General Relativity within our framework. The physical theory tells us that two groups act on the local fields \(E\) and \(A\), encoding the symmetries of the theory. These are the group of vertical automorphisms \(\mathcal{G} = \mathrm{Gau}(\SP)\) and the diffeomorphism group \(\Diff\).\\

The geometrical description of the Ashtekar-Barbero-Immirzi formulation is given by a pair \(\tau\) and \(\omega\) of \(\mathfrak{su}(2)\)-valued differential forms on \(\SP\). However, in the physical description, we work with fields on \(\Sigma\). Since \(\SP\) is trivial over \(\Sigma\), it is possible to establish a natural correspondence between the geometric data and the local fields by fixing a global trivialization. Let $s$ be the global section associated with the trivialization, in this case, $\tau$ is mapped to an electric field $E=s^*\tau\in\Omega^2(\Sigma)\tensor \su{2}^*$ and $\omega$ to an Ashtekar connection $A=s^*\omega\in\Omega^1(\Sigma)\tensor \su{2}$. The group of gauge transformations has a convenient isomorphism $\mathcal{G}\cong C^{\infty}(\Sigma,SU(2))$, and the group action can be deduced by the pullback. Let $f\in \mathcal{G}$ be identified by $g:\Sigma\to SU(2)$, then $f^*\tau\mapsto\Ad_{g^{-1}}^*E$ and $f^*\omega\mapsto \Ad_{g^{-1}}A+g^{-1}dg$. While $\Diff$ acts via pullback of the $k$-form part of the local fields.\\

This is the local version of saying that $\mathrm{Aut}(\SP)$ is the symmetry group of the theory. Indeed, it is well known that, since $\SP$ is trivial, the following short sequence holds \cite{Abbati89}
\[1\to\mathcal{G}\to\mathrm{Aut}(\SP)\to\Diff\to1.\]
Moreover, there exists an homomorphism from $\Diff$ to $\mathrm{Aut}(\SP)$ which maps, in a fixed trivialization, $\varphi\mapsto \varphi\times \mathrm{id}_{SU(2)}$. That map is projected by the second homomorphism into the identity on $\Diff$. This condition is necessary and sufficient for $\mathrm{Aut}(\SP)$ to be isomorphic to $\mathcal{G}\rtimes\Diff$. Notice that $\mathrm{Aut}(\SP)$ is the group of isomorphisms in the category of principal bundles, providing a natural parallelism between physical symmetry and geometric equivalence. \\

A deeper analysis of these symmetry groups is provided in the context of symplectic geometry in the next Section.

\section{A note on phase space}
In this Section, we aim to characterize the phase space of the theory in terms of symplectic geometry. We are going to provide a symplectic form and show that the Gauss and Vector constraint can be interpreted as momentum maps on this symplectic manifold. The construction of the phase space emulates the procedure proposed in \cite{thiemann_2007}, namely we will consider as elements of the phase space the couples of local fields $(E,A)$, which in our framework means to fix a global trivialization (equivalently choose a section) for $\SP$. However, the generalization to vector-valued sections and connections is quite straightforward.\\

Let $\Sigma$ be a spinnable closed 3-manifold. From what we have seen before, we can collect the data of the metric in an $\mathrm{ad}^*\SP$-valued 2-form $\boldsymbol{E}$. In a fixed trivialization, it is represented by a 2-form $E=E^a_i\epsilon_{abc} dx^a\wedge dx^b \mathfrak{t}^i$ with values in $\su{2}^*$. Such a 2-form has to be nondegenerate, namely $\det(E^a_i)$ is never zero. Thus, the configuration space is $\mathcal{E}\doteq\{E\in \Omega^2(\Sigma)\tensor\su{2}^*\ |\ E \textrm{ is non degenerate}\}$. The space $\mathcal{E}$ is a Fréchet manifold. Indeed, it is an open subset of $\Omega^2(\Sigma)\tensor\su{2}^*$, which is a Fréchet topological vector space. The Fréchet space structure is inherited, via pullback map of the chosen section, by the one of $\Omega^2(\Sigma,\mathrm{ad}^*\SP)$, which is a Fréchet space for a fixed choice of a triple composed by an auxiliary metric on $\Sigma$, a Riemannian product on $\mathrm{ad}^*\SP$, and a auxiliary connection on $\SP$. By compactness of $\Sigma$, the Fréchet space structure does not depend on the choice of the triple \cite{Bar_2015}. The family of seminorms on $\Omega^2(\Sigma,\mathrm{ad}^*\SP)$ is given by
\begin{equation}
    ||\boldsymbol{E}||_m=\max_{l\in\{0,\dots,m\}}\max_{x\in\Sigma}|\nabla^l \boldsymbol{E}(x)|,\ m\in\mathbb{N}_0,
\end{equation}
where $|\cdot|$ is the norm induced on $\T^*\Sigma^{\tensor^l}\tensor\Lambda^2\T^*\Sigma\tensor\mathrm{ad}^*\SP$ by the auxiliary metric and the Riemannian product, and $\nabla$ is the covariant derivative induced on $\T^*\Sigma^{\tensor^l}\tensor\Lambda^2\T^*\Sigma\tensor\mathrm{ad}^*\SP$ by the Levi-Civita connection and the chosen auxiliary connection.
More concretely, we can consider a collection of seminorms on $\Omega^2(\Sigma)\tensor\su{2}^*$ as given by an auxiliary metric, the scalar product on $\su{2}$ defined in \eqref{defE}, and a flat connection, namely
\begin{align*}
    &||E||_0=\max_{x\in\Sigma}|E_a(x)\mathfrak{t}^a|,\\
    &||E||_1=\max\left\{||E||_0,\,\max_{x\in\Sigma}\left|\left(\nabla^{\Lambda^2} E_a(x)\right)\mathfrak{t}^a\right|\right\},\\
    &||E||_2=\max\left\{||E||_1,\,\max_{x\in\Sigma}\left|\nabla^{\Lambda^1\tensor\Lambda^2}\left(\nabla^{\Lambda^2} E_a(x)\right)\mathfrak{t}^a\right|\right\},\\
    &||E||_3=\max\left\{||E||_2,\,\max_{x\in\Sigma}\left|\left(\nabla^{\Lambda^1\tensor\Lambda^1\tensor\Lambda^2}\nabla^{\Lambda^1\tensor\Lambda^2}\nabla^{\Lambda^2} E_a(x)\right)\mathfrak{t}^a\right|\right\},\\
    &\vdots
\end{align*}
where $\nabla^{\Lambda^i}$ is the Levi-Civita connection of the auxiliary metric induced on the exterior power $\Lambda^i\T^*\Sigma$ and $|\cdot|$ is the norm induced by the auxiliary metric on $\T^*\Sigma^{\tensor^l}\tensor\Lambda^2\T^*\Sigma$ and the scalar product $tr$ defined before.
As a consequence, the tangent space on each point is isomorphic to $\Omega^2(\Sigma)\tensor\su{2}^*$. In particular, the tangent bundle is trivial and $\mathcal{E}$ is parallelizable, i.e. $\T \mathcal{E}\cong \mathcal{E}\times \left(\Omega^2(\Sigma)\tensor\su{2}^*\right)$. Notice that $\mathcal{E}$ has two connected components associated with the two possible orientations of $\Sigma$, namely the sign of $\det(E^a_i)$.\\\\

Each $E$ is associated to a metric, constructed as in Sec.\ref{SubMetr}. Given this metric and fixing a spin structure, we can define as in Sec.\ref{SecGauge} the space of Ashtekar connections $\mathcal{A}_E$. These spaces do not actually depend on the metric, and they are canonically the space of $SU(2)$-connections $\mathcal{A}$.\\
The space of connections $\mathcal{A}$ is an affine space associated with the vector space $\Omega^1(\Sigma,\mathrm{ad}\SP)$, then it is a Fréchet manifold. Fixing a global trivialization, it is in one-to-one correspondence with $\Omega^1(\Sigma)\tensor\su{2}$. From which we are going to consider as phase space $\mathcal{P}$ the fiber bundle with base manifold $\mathcal{E}$ and fiber $\mathcal{A}$ (the fiber over each point is $\mathcal{A}_E$). Such a bundle admits a global section, assigns to each $E$ the corresponding spin connection $\Gamma(E)$ , hence it is a trivial bundle $\mathcal{P}\cong\mathcal{E}\times\mathcal{A}$.\\
The tangent bundle $\T\mathcal{P}$ can be easily characterized $\T\mathcal{P}\cong\T\mathcal{E}\times\T\mathcal{A}\cong\mathcal{E}\times \left(\Omega^2(\Sigma)\tensor\su{2}^*\right)\times\mathcal{A}\times \left(\Omega^1(\Sigma)\tensor\su{2}\right)$. Hence, on each point, $\T_{(E,A)}\mathcal{P}\cong \T_E\mathcal{E}\oplus \T_A\mathcal{A}\cong\left(\Omega^2(\Sigma)\tensor\su{2}^*\right)\oplus\left(\Omega^1(\Sigma)\tensor\su{2}\right)$.
Notice that there is a weakly nondegenerate pairing between $\Omega^2(\Sigma)\tensor\su{2}^*$ and $\Omega^1(\Sigma)\tensor\su{2}$:
\begin{align*}
    \left(\Omega^2(\Sigma)\tensor\su{2}^*\right)\times\left(\Omega^1(\Sigma)\tensor\su{2}\right)\to \R;\\
    (\nu,\alpha)\mapsto\int_{\Sigma}tr( \nu\wedge \alpha).
\end{align*}
From which, we can define the canonical symplectic form $\omega$ \cite{Marsden1999} on the tangent space of a point $(E,A)\in\mathcal{P}$ by
\begin{align}
\nonumber
    \omega_{(E,A)}:\T_{(E,A)}\mathcal{P}\times\T_{(E,A)}\mathcal{P}&\to \R\\
    ((\nu_1,\alpha_1),(\nu_2,\alpha_2))&\mapsto \int_{\Sigma}tr(\nu_1\wedge \alpha_2)-\int_{\Sigma}tr(\nu_2\wedge \alpha_1)
\end{align}
The symplectic form is defined modulo a constant real factor $\beta$ which represents the Barbero-Immirzi parameter. Actually, it can be associated with the symplectic potential
\begin{align*}
    \theta_{(E,A)}:\T_{(E,A)}\mathcal{P}&\to \R\\
     (\nu,\alpha)&\mapsto \int_{\Sigma}tr(E\wedge \alpha)
\end{align*}
Notice that it is the same symplectic structure of Yang-Mills theory \cite{Segal_1981,Sniatycki_1999}. Introducing the exterior derivative $\delta$ on $\mathcal{P}$, those objects can be written, with a slight abuse of notation, as $\omega=\int tr(\delta E\barwedge\delta A)$ and $\theta=\int tr(E\wedge\delta A)$, where $\barwedge$ represents the wedge product between $k$-forms on $\Sigma$ and between $l$-forms of $\mathcal{P}$.\\\\
The constraints are well-defined on this phase space. The Gauss constraint $G$ and the Vector constraint $\mathrm{V}$ can be regarded as functions on the phase space with values in suitable infinite-dimensional Fréchet Lie algebras. These functions are globally defined due to the smoothness of the fields and the compactness of $\Sigma$. Moreover, they have an interpretation in terms of momentum maps, and we will implement them within the framework of symplectic geometry via symplectic reduction. Note that if we were to consider the space of $L^2$ connections instead, we would also need to analyze the domain of the constraints, likely involving some type of Sobolev space.

\subsection{Gauge transformations and momentum maps}
The infinite-dimensional group of gauge transformations $\mathcal{G}\cong C^{\infty}(\Sigma,SU(2))$ acts on this phase space. The group has a natural right action on $\mathcal{P}$:
\begin{align}
\nonumber
    \mathcal{G}\times \mathcal{P}&\to\mathcal{P};\\
    (g,(E,A))&\mapsto g.(E,A)\doteq(\Ad^*_{g^{-1}}E,\Ad_{g^{-1}}A+g^{-1}dg)
\end{align}
The action preserves the symplectic form $\omega$, i.e. it acts via symplectomorphisms. 
\begin{align*}
    (g.^*\omega)_{(E,A)}((\nu_1,\alpha_1),(\nu_2,\alpha_2))=\omega_{g.(E,A)}(\delta g.(\nu_1,\alpha_1),\delta g.(\nu_2,\alpha_2))\\
    =\omega_{g.(E,A)}((\Ad^*_{g^{-1}}\nu_1,\Ad_{g^{-1}}\alpha_1),(\Ad^*_{g^{-1}}\nu_2,\Ad_{g^{-1}}\alpha_2))\\
    =\int_{\Sigma}tr(\Ad^*_{g^{-1}}\nu_1\wedge \Ad_{g^{-1}}\alpha_2)-\int_{\Sigma}tr(\Ad^*_{g^{-1}}\nu_2\wedge \Ad_{g^{-1}}\alpha_1)\\
    =\int_{\Sigma}tr(\nu_1\wedge \alpha_2)-\int_{\Sigma}tr(\nu_2\wedge \alpha_1)
\end{align*}
Moreover, it is possible to define the momentum map for this action as follows: 
\begin{align}
\nonumber
    \mu_G:\mathcal{P}&\to \Omega^3(\Sigma)\tensor\su{2}^*;\\
    (E,A)&\mapsto G(E,A)=d_A E
\end{align}
The map $\mu_G$ is equivariant with respect to the $\mathcal{G}$-action: $G(g.(E,A))=\Ad^*_{g^{-1}}G(E,A)$. The coupling with $Lie(\mathcal{G})\cong\Omega^0(\Sigma)\tensor \su{2}$ is given by
\begin{align*}
    \left(\Omega^3(\Sigma)\tensor\su{2}^*\right)\times\left(\Omega^0(\Sigma)\tensor\su{2}\right)\to \R;\\
    (\chi,\xi)\mapsto\bra\chi,\xi\ket\doteq\int_{\Sigma}tr( \chi \xi)
\end{align*}
We can now prove the momentum map property. Considering the vector field on $\mathcal{P}$ defined by \[\varrho(\xi)_{(E,A)}\doteq\frac{d}{dt}\Big|_0\exp(t\xi).(E,A)=\left(-\ad^*(\xi)E,d\xi+\ad(A)\xi\right),\]
we must show $\iota_{\varrho(\xi)}\omega=\delta\bra\mu_G,\xi\ket$. Let us compute first the interior product with the symplectic form
\begin{align*}
    &(\iota_{\varrho(\xi)}\omega)(\nu,\alpha)=\omega\left((-\ad^*(\xi)E,d\xi+\ad(A)\xi),(\nu,\alpha)\right)\\
    &=-\int_{\Sigma}tr(\ad^*(\xi)E\wedge \alpha)-\int_{\Sigma}tr(\nu\wedge (d\xi+\ad(A)\xi)).
\end{align*}
Then, we compute the differential of the coupling with a slight abuse of notation
\begin{align*}
    &\delta\bra\mu_G,\xi\ket=\delta\int_{\Sigma}tr( G(E,A) \xi)=\int_{\Sigma}tr\left( (d\delta E+\ad^*(\delta A)\wedge E+\ad^*(A)\wedge \delta E) \xi\right)\\
    &=-\int_{\Sigma}tr(\delta E\wedge d\xi)-\int_{\Sigma}tr( \ad^*(\xi)E\wedge \delta A)-\int_{\Sigma}tr( \delta E\wedge \ad(A)\xi).
\end{align*}
Thus, the $\mathcal{G}$-action on $\mathcal{P}$ is Hamiltonian and we can consider the symplectic reduction $\mathcal{P}\sslash\mathcal{G}=\mu^{-1}_G(0)/\mathcal{G}$. However, this action is not free because there exists a non-trivial center. We can slightly modify the definition of gauge transformation considering $\mathcal{G}\doteq C^{\infty}(\Sigma,\Ad SU(2))$. This choice does not modify the Lie algebra and the previous considerations, moreover, the action is free on $\mathcal{E}$ and so on $\mathcal{P}$ even if $\mathcal{A}$ admits reducible connections \cite{Nash_1991}.

\subsection{Diffeomorphisms group and ADM phase space}
A second group acts via symplectomorphism on $\mathcal{P}$, it is $\Diff$.\\
The action is $\varphi.(E,A)=(\varphi^*E,\varphi^*A)$, it passes to the quotient and the symplectic form is invariant under it:
\begin{align*}
    (\varphi.^*\omega)_{(E,A)}((\nu_1,\alpha_1),(\nu_2,\alpha_2))=\omega_{\varphi.(E,A)}(\delta \varphi.(\nu_1,\alpha_1),\delta \varphi.(\nu_2,\alpha_2))\\
    =\omega_{\varphi.(E,A)}((\varphi^*\nu_1,\varphi^*\alpha_1),(\varphi^*\nu_2,\varphi^*\alpha_2))\\
    =\int_{\Sigma}tr(\varphi^*\nu_1\wedge \varphi^*\alpha_2)-\int_{\Sigma}tr(\varphi^*\nu_2\wedge \varphi^*\alpha_1)\\
    =\int_{\Sigma}\varphi^*tr(\nu_1\wedge \alpha_2)-\int_{\Sigma}\varphi^*tr(\nu_2\wedge \alpha_1)\\
    =\int_{\varphi(\Sigma)}tr(\nu_1\wedge \alpha_2)-\int_{\varphi(\Sigma)}tr(\nu_2\wedge \alpha_1)\\
    =\int_{\Sigma}tr(\nu_1\wedge \alpha_2)-\int_{\Sigma}tr(\nu_2\wedge \alpha_1).
\end{align*}
The map given by $(E, A)\mapsto \mathrm{V}(E, A)=tr\left(e(E)\tensor\ad^*(F(A))\wedge e(E)\right)$ is invariant under the action of $\mathcal{G}$, hence it defines a well-defined map on the quotient space. Moreover, its range is in $\Omega^3(\Sigma)\tensor\Omega^1(\Sigma)$ and that map is $\Diff$-equivariant $\mathrm{V}(\varphi^*E,\varphi^*A)=\varphi^*\mathrm{V}(E,A)$. Its projection to the quotient is the momentum map associated with $\Diff$, which is the supermomentum constraint.\\
Using the procedure explained in Sec.\ref{SubMetr}, it is possible to associate a metric $q(E)$ to $E$, which is the same for the entire equivalent class. In the same proof, the cotriad $e(E)$ reconstructed from $E$ is also defined. Moreover, we can define the function $\Gamma(E)$ that is the Levi-Civita connection of the metric $q(E)$ with respect to the triad reconstructed by $E$. Such a function is a homogeneous function of $E$ of degree zero \cite{thiemann_2007}. Finally, we can associate a tensor $\mathrm{K}(E,A)=tr\left(e(E)\tensor (A-\Gamma(E))\right)$ that it constant along the $\mathcal{G}$-orbits in $\mathcal{P}$ and symmetric on $\mu^{-1}_G(0)$. Changing the orientation is equivalent to consider the transformation $E\mapsto-E$, we can notice that $q(E)=q(-E)$ and $\mathrm{K}(E,A)=-\mathrm{K}(-E,A)$. Hence, the equivalence classes $[E,A]$ in $\mathcal{P}\sslash\mathcal{G}$ are identified by couples $(q(E),\mathrm{K}(E,A))$. Therefore, the reduced symplectic manifold $\mathcal{P}\sslash\mathcal{G}$ is a fiber bundle with base manifold the space of Riemannian metrics $\mathit{Met}(\Sigma)$ and as fiber the vector space $C^{\infty}(\Sigma,\mathrm{Sym}(\T\Sigma))$ of symmetric tensors. Since $\mathit{Met}(\Sigma)$ is contractible, the fiber bundle is trivial.\\
The action of the Diffeomorphism group is induced by the one on $\mathcal{P}$ and reads simply $(q,\mathrm{K})\mapsto (\varphi^*q,\varphi^*\mathrm{K})$. It acts via symplectomorphism and has a momentum map that is the supermomentum constraint. 
This is induced on the reduced manifold by the projection of the Vector constraint, while the Scalar constraint is projected into the superhamiltonian one. Hence, this phase space is isomorphic to the ADM one \cite{ashtekar86,ashtekar87}.

\subsection{Full group of symmetries}
We can consider the action of the full group $\mathfrak{G}=\mathcal{G}\rtimes\Diff$, where the group product is
\begin{align*}
    (g_1,\varphi_1)\cdot(g_2,\varphi_2)=(g_1\cdot( g_2\circ\varphi_1^{-1}),\varphi_1\circ\varphi_2)
\end{align*}
The action of the couple $(g,\varphi)$ on $\mathcal{P}$ is 
\begin{equation}
    (g,\varphi).(E,A)=(\Ad^*_{g^{-1}}\varphi^*E,\Ad_{g^{-1}}\varphi^*A+g^{-1}dg).
\end{equation}
The Lie algebra of this group is isomorphic as vector space to $Lie(\mathcal{G})\oplus Lie(\Diff)$. Following some basic calculations about semidirect products of groups, we find the adjoint action on $(\xi, X)$, where $\xi\in C^{\infty}(\Sigma,\su{2})\cong Lie(\mathcal{G})$ and $X\in\mathfrak{X}(\Sigma)\cong Lie(\Diff)$:
\begin{equation}
    \label{full-adj}
    \Ad_{(g,\varphi)}(\xi,X)=(\Ad_g \xi\circ\varphi^{-1}-gdg^{-1}(\varphi_*X),\varphi_*X).
\end{equation}
From which, we can find the coadjoint representation on $\left(\Omega^3(\Sigma)\tensor\su{2}^*\right)\oplus\left(\Omega^3(\Sigma)\tensor\Omega^1(\Sigma)\right)$ by imposing 
\begin{align*}
    &\bra \Ad^*_{(g,\varphi)}(\chi,\lambda),(\xi,X)\ket=\bra (\chi,\lambda),\Ad_{(g,\varphi)^{-1}}(\xi,X)\ket\\
    &=\int_{\Sigma}tr\left(\chi\tensor(\Ad_{g^{-1}\circ\varphi}\xi\circ\varphi-(g^{-1}\circ \varphi) d(g\circ\varphi)(\varphi^{-1}_*X))\right)+\int_{\Sigma}\lambda(\varphi_*^{-1}X)\\
    &=\int_{\Sigma}tr\left((\Ad^*_{g}(\varphi^{-1})^*\chi)\tensor \xi\right)- \int_{\Sigma}tr\left((\varphi^{-1})^*\chi\tensor g^{-1}dg(X)\right)+\int_{\Sigma}(\varphi^{-1})^*\lambda(X).
\end{align*}
Then, the coadjoint action reads
\begin{equation}
\label{full-coadj}
    \Ad^*_{(g,\varphi)}(\chi,\lambda)=(\Ad^*_g (\varphi^{-1})^*\chi,(\varphi^{-1})^*\lambda+tr((\varphi^{-1})^*\chi\tensor gdg^{-1}).
\end{equation}
For completeness, we can compute the adjoint action of the algebra on itself, namely the Lie bracket:
\begin{equation}
    \ad_{(\xi_1,X_1)}(\xi_2,X_2)=(\ad_{\xi_1}\xi_2+d\xi_1(X_2)-d\xi_2(X_1),-\mathcal{L}_{X_1}X_2).
\end{equation}
We can now discuss the momentum map of this full group, which acts via symplectomorphisms.\\
It is known that the correct constraint that implements diffeomorphisms is a deformation of (but equivalent to) the Vector constraint. In our case, the Diffeomorphism constraint reads
\begin{equation}
    \mathrm{D}(E,A)=\mathrm{V}(E,A)+tr(G(E,A)\tensor A)
\end{equation}
The momentum map is defined by:
\begin{align}
    \nonumber
    \mu:\mathcal{P}\to &\left(\Omega^3(\Sigma)\tensor\su{2}^*\right)\oplus\left(\Omega^3(\Sigma)\tensor\Omega^1(\Sigma)\right);\\
    (E,A)\mapsto &\mu(E,A)=(G(E,A),\mathrm{D}(E,A)).
\end{align}
This map is $\mathfrak{G}$-equivariant with respect to the right action
\begin{align*}
    \mu\circ(g,\varphi).(E,A)&=(\Ad^*_{g^{-1}}\varphi^*G(E,A),\varphi^*\mathrm{D}(E,A)+tr(\varphi^*G(E,A)\tensor g^{-1}dg))\\
    &=\Ad^*_{(g^{-1},\varphi^{-1})}\mu(E,A).
\end{align*}
From the literature \cite{thiemann_2007}, with the same slight abuse of notation as before, we can find that 
\begin{align*}
    \delta\bra\mu_D,X\ket=\int_{\Sigma}tr(\mathcal{L}_X E\wedge\delta A)-\int_{\Sigma}tr(\delta E\wedge\mathcal{L}_X A).
\end{align*}
Considering now the vector field $\varrho(X)_{(E, A)}\doteq\left(\mathcal{L}_X E,\mathcal{L}_X A\right)$ on $\mathcal{P}$ that generates the Diffeomorphism transformations, the interior product matches the differential of the pairing
\begin{align*}
    &(\iota_{\varrho(X)}\omega)(\nu,\alpha)=\omega\left(\mathcal{L}_X E,\mathcal{L}_X A),(\nu,\alpha)\right)\\
    &=\int_{\Sigma}tr(\mathcal{L}_X E\wedge \alpha)-\int_{\Sigma}tr(\nu\wedge \mathcal{L}_X A).
\end{align*}
The previous computations show that the Diffeomorphism constraint $\mathrm{D}(E,A)$ is the momentum map for the $\Diff$-action and has the correct transformation behavior under $\mathcal{G}$-action in order to $\mu$ be the momentum map for the $\mathfrak{G}$-action on $\mathcal{P}$
\[\delta\bra\mu,(\xi,X)\ket=\iota_{\varrho(\xi,X)}\omega.\]
Concerning the usual physical notation, the Hamiltonian functions of the respective group actions, namely the natural pairing between momentum maps and Lie algebra elements, are the so-called smeared constraints.

\section{Conclusion}
The Ashtekar-Barbero-Immirzi formulation of General Relativity admits a geometric framework that is analogous to an \( SU(2) \) gauge theory. In this formulation, the configuration variables are interpreted as a connection on a principal \( SU(2) \)-bundle over \( \Sigma \), similar to the case in \( SU(2) \) Yang-Mills theory. However, to establish this interpretation, it is necessary to introduce an additional structure: the spin structure. The spin structure represents the main conceptual difference from Yang-Mills theory. It gives the Ashtekar connection a geometric meaning, treating it on the same level as the Levi-Civita connection, which is crucial for the reconstruction and the geometric link with the ADM formulation.\\
Within this framework, we provide a coordinate-free definition of the Gauss, Vector, and Scalar constraints, which take the form of vector-bundle-valued 3-forms and have a geometric interpretation in terms of the exterior covariant derivative.\\
In this formulation, we can implement the physical symmetries, demonstrating that they are represented by the group of automorphisms of the bundle. Additionally, we can analyze the phase space, which is a Fréchet manifold, within the context of symplectic geometry. Within our framework, we recover the well-known relationship between Ashtekar variables and the ADM phase space through symplectic reduction. Furthermore, we prove that the action of the group of automorphisms is a Hamiltonian action with an equivariant momentum map, given by the pair of Gauss and Diffeomorphism constraints, expressed as vector-bundle-valued 3-forms.\\

We hope that this formalization of the classical theory will enable a mathematically rigorous approach to quantum theory. In this regard, a recent study attempts to implement this formalism within a cosmological framework, exploring its implications for the quantum cosmological sector of the full quantum theory \cite{Bruno_2024}. We believe that the mathematical physics community has the potential to make meaningful and significant contributions to the understanding and advancement of this theory, paving the way for new insights and developments in the field.

\section*{Acknowledgment}
The author would like to express his gratitude to Gianluca Panati and Domenico Fiorenza for their guidance and interest during the composition of this manuscript. The author also extends his sincere thanks to Giovanni Montani, Fabio M. Mele, Eugenia Colafranceschi, and Lorenzo Boldorini for their valuable contributions to fruitful discussions.

\appendix
\section{Vierbein and spin structure on spacetime}
\label{AppTetrads}
The bundle structure is inherently present on the spacetime $\M$ and becomes evident through the vierbein formalism. In this approach, the spacetime is equipped with a local frame at each point, which can be interpreted as a section of a principal bundle.\\
Suppose $\M$ is orientable, the orthonormal frame bundle $P^{SO}(\M)$ of $\M$ is well-defined. Because the metric $g$ is Lorentzian, the orthonormal frame bundle is a principal $SO(1,3)$-bundle. An element of this bundle in the fiber over $p\in\M$ is an orientation preserving linear isometry $h:\R^{1,3}\to \T_p\M$.\\
Hence, the vierbein is the data of a section in $P^{SO}(\M)$. It defines a set of orthonormal vectors $\{e_{\alpha}\}$, i.e. $g(e_{\alpha},e_{\beta})=\eta_{\alpha\beta}$, where $\eta_{\alpha\beta}=\mathrm{diag}(-1,1,1,1)$.\\
In this setting, the spin connection $\omega^{\alpha\beta}_{\mu}$, defined in Eq.\ref{constraint}, is the connection 1-form of the Levi-Civita connection in the orthonormal basis:
\[\nabla^{LC}e_{\alpha}=\omega^{\beta}_{\alpha}e_{\beta}.\]
Thus, Eq.(\ref{constraint}) is interpreted as the covariant derivative of the Casimir element of the endomorphisms on the tangent space: $\nabla^{LC}\operatorname{Id}_{\T\M}\equiv 0$, that is automatically satisfied in this framework. Hence, the constraint (\ref{constraint}) allows us to consider this fiber bundle construction.\\\\
The existence of spinor fields on spacetime is a physical fact, necessary for the description of the elementary particles. Hence, the existence of a spin structure on $\M$ is a proper hypothesis. Given a spacelike hypersurface in $\M$, it can be equipped with a unique spin structure induced by the one on $\M$. Let $S$ be a spacelike hypersurface, there exists a canonical embedding $\iota:P^{SO}(S)\imm~P^{SO}(\M)$. Let $n$ be the normal versor field along $S$, this embedding map is given by $\iota:(h:\R^3\to \T_pS)\mapsto (h':\R^{1,3}\to \T_p\M$), where $h'(0,v^1,v^2,v^3)=h(v^1,v^2,v^3)$ and $h'(1,0,0,0)=n(p)$. This embedding is compatible with the canonical embedding $SO(3)\imm SO(1,3)$, so their right action on their respective orthonormal frame bundles commute with the embedding. Let $(P^{Spin}(\M),\Bar{\varrho})$ be the spin structure on $\M$, the spin structure on $S$ is $(P^{Spin}(S)\doteq\Bar{\varrho}^{-1}(P^{SO}(S)),\Bar{\varrho})$ \cite{Bar}.\\\\
In the case of a globally hyperbolic spacetime $\M\cong\R\times\Sigma$, for the above discussion every Cauchy surface $f^{-1}(t)$ admits a spin structure. Here, $f$ is the surjective function of Geroch's theorem.\\
Notice that the vierbein defined in (\ref{vierred}) corresponds to a choice of an element \(h'\) of \(P^{SO}(\mathcal{M})\), as defined above. Specifically, let \(\mathcal{M} = \mathbb{R} \times \Sigma\), with a chosen coordinate system as in Sec.~\ref{SecADM}, and a dreibein \(e\) over the Cauchy surface \(\Sigma_t\). The vierbein \(h'\) defined by the embedding \(\iota\circ e\) has a matrix form that is exactly the inverse of the matrix in (\ref{vierred}). Indeed, let us consider $\mathfrak{e}_{\alpha}$, with $\alpha=0,1,2,3$, as the canonical basis of $\R^{1,3}$. Hence, the vierbein satisfies $\iota (e(\mathfrak{e}_0))=n=\frac{1}{N}\partial_0-\frac{N_a}{N}\partial_a=e^{\mu}_{0}\partial_\mu$, and $\iota(e(\mathfrak{e}_i))=e(\mathfrak{e}_i)=e^a_i\partial_a$.
\\\\
We can also provide a topological algebraic argument to show that if $\M=\R\times \Sigma$ admits a spin structure, then $\Sigma$ admits one too. We will check this by computing the Stiefel-Whitney classes \cite{LawsonSpin}. Notice that $\T\M\cong\T\R\times\T\Sigma\cong \mathrm{pr}_1^*\T\R\oplus\mathrm{pr}_2^*\T\Sigma$. Here, $\mathrm{pr}_1:\M\to\R$ and $\mathrm{pr}_2:\M\to\Sigma$ are the projectors on the first and second factor, respectively. Using the naturality of Stiefel-Whitney classes and the Whitney sum formula, and recalling that $w_0\equiv 1,$ we get
\begin{align*}
    & w_1(\T\M)=w_1(\mathrm{pr}_1^*\T\R\oplus\mathrm{pr}_2^*\T\Sigma)=w_1(\T\Sigma),\\
    & w_2(\T\M)=w_2(\mathrm{pr}_1^*\T\R\oplus\mathrm{pr}_2^*\T\Sigma)=w_2(\T\Sigma).
\end{align*}
Hence, if $\M$ is spinnable, then $w_1(\T\Sigma) =  w_1(\T\M) = 0$ and $w_2(\T\Sigma) = w_2(\T\M) = 0$, meaning that $\Sigma$ is also spinnable. Thus, the hypothesis of the existence of a spin structure on $\Sigma$ is determined by the same condition on the globally hyperbolic spacetime $\M$.

\bibliographystyle{elsarticle-num} 
\bibliography{biblio}

\end{document}